\documentclass[10pt,twocolumn,letterpaper]{article}

\usepackage{iccv}              

\usepackage{amsmath,amsfonts,bm}

\def\eqref#1{equation~\ref{#1}}

\def\1{\bm{1}}

\def\rvc{{\mathbf{c}}}
\def\rvd{{\mathbf{d}}}

\def\rvf{{\mathbf{f}}}

\def\rvl{{\mathbf{l}}}

\def\rvq{{\mathbf{q}}}
\def\rvr{{\mathbf{r}}}
\def\rvs{{\mathbf{s}}}

\def\rvx{{\mathbf{x}}}

\def\rmO{{\mathbf{O}}}

\DeclareMathAlphabet{\mathsfit}{\encodingdefault}{\sfdefault}{m}{sl}
\SetMathAlphabet{\mathsfit}{bold}{\encodingdefault}{\sfdefault}{bx}{n}

\def\sR{{\mathbb{R}}}

\definecolor{iccvblue}{rgb}{0.21,0.49,0.74}
\usepackage[pagebackref,breaklinks,colorlinks,allcolors=iccvblue]{hyperref}

\def\paperID{9755} 
\def\confName{ICCV}
\def\confYear{2025}

\title{SD-GS: Structured Deformable 3D Gaussians for Efficient Dynamic Scene Reconstruction}

\author{
\begin{tabular}{c}
Wei Yao$^{1\star}$, Shuzhao Xie$^{1\star}$, Letian Li$^1$, Weixiang Zhang$^1$, \\
Zhixin Lai$^2$, Shiqi Dai$^3$, Ke Zhang$^4$, Zhi Wang$^{1\dagger}$ \\
\textnormal{\normalsize
$^1$SIGS, Tsinghua University \quad
$^2$Google \quad
$^3$Department of CST, Tsinghua University \quad
$^4$Soochow University}\\
\texttt{\normalsize \{yaow21, xsz24, lilt24, zhang-wx22\}@mails.tsinghua.edu.cn, zhixinlai@google.com,} \\
\texttt{\normalsize daisq99@gmail.com, kzhang19@suda.edu.cn, wangzhi@sz.tsinghua.edu.cn}
\end{tabular}
}

\begin{document}

\twocolumn[{%
\renewcommand\twocolumn[1][]{#1}%
\maketitle
\begin{center}
\centering
\vspace{-10pt}
\includegraphics[width=1\linewidth]{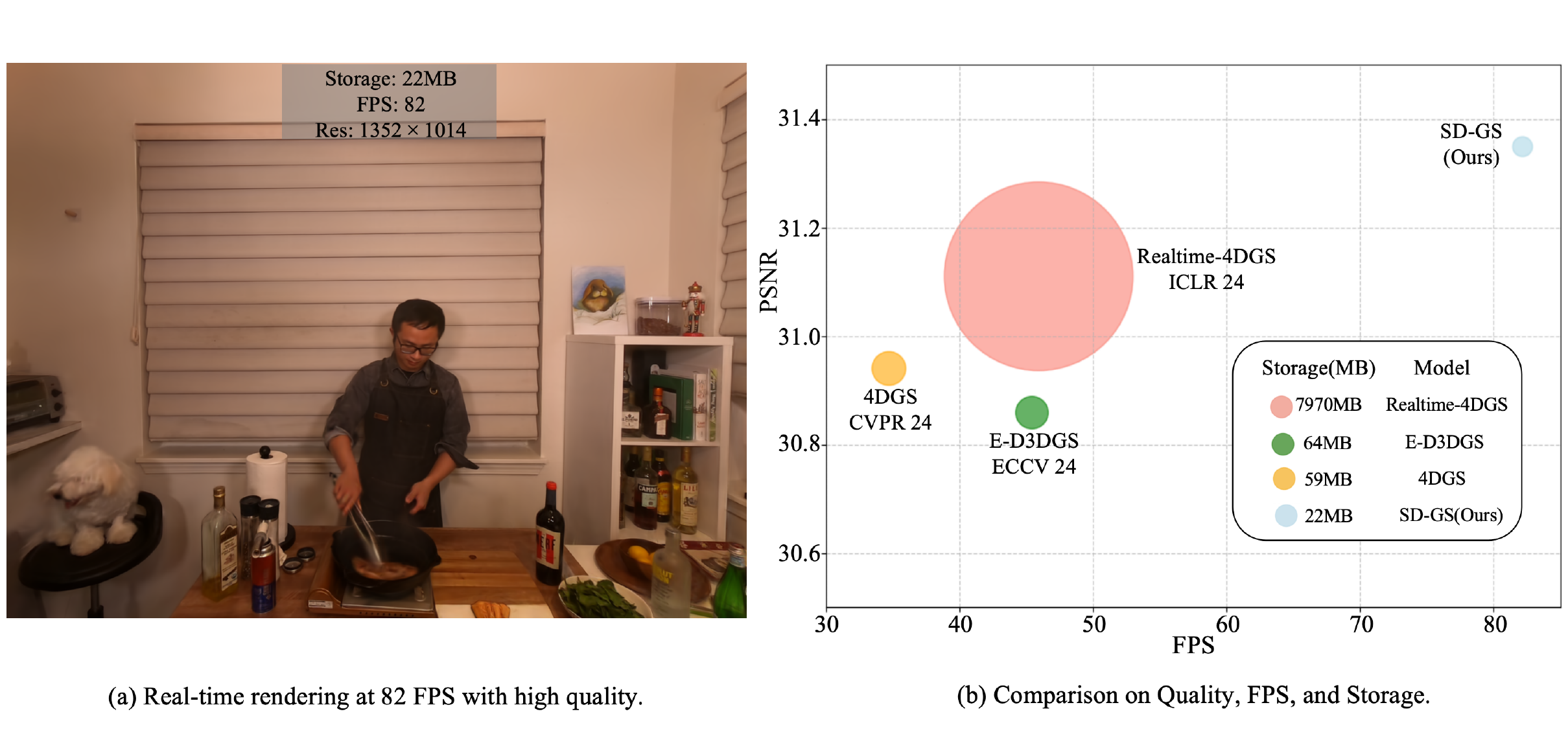}
\label{fig:1}
\vspace{-18pt}
\captionof{figure}{Our method successfully achieves photorealistic quality and high resolution rendering in real time while maintaining a compact model size. (a) Our approach can be rendered at high resolution with 82 FPS on an Nvidia RTX 3090 GPU. (b) Quantitative comparisons of rendering quality, speed, and storage requirements with several state-of-the-art baselines on the N3DV Dataset.}
\end{center}%
}]

\def\thefootnote{$\star$}\footnotetext{Equal contribution.}
\def\thefootnote{$\dagger$}\footnotetext{Corresponding author.}

\begin{abstract}
Current 4D Gaussian frameworks for dynamic scene reconstruction deliver impressive visual fidelity and rendering speed, however, the inherent trade-off between storage costs and the ability to characterize complex physical motions significantly limits the practical application of these methods.
To tackle these problems, we propose SD-GS, a compact and efficient dynamic Gaussian splatting framework for complex dynamic scene reconstruction, featuring two key contributions.
%
First, we introduce a deformable anchor grid, a hierarchical and memory-efficient scene representation where each anchor point derives multiple 3D Gaussians in its local spatiotemporal region and serves as the geometric backbone of the 3D scene.
%
Second, to enhance modeling capability for complex motions, we present a deformation-aware densification strategy that adaptively grows anchors in under-reconstructed high-dynamic regions while reducing redundancy in static areas, achieving superior visual quality with fewer anchors.
Experimental results demonstrate that, compared to state-of-the-art methods, SD-GS achieves an average of 60\% reduction in model size and an average of 100\% improvement in FPS, significantly enhancing computational efficiency while maintaining or even surpassing visual quality.


\end{abstract}    
\section{Introduction}
\label{sec:intro}
Dynamic scene reconstruction from multi-view videos is an important task in 3D computer vision, with tremendous applications in AR, VR, and 3D content creation~\cite{tang2023dreamgaussian}.
While Neural Radiance Fields (NeRFs) \citep{li2022neural3d,wang2023mixed,attal2023hyperreel,pumarola2021d, park2021hypernerf} have made notable progress in dynamic scene reconstruction, 3D Gaussian Splatting (3DGS)-based methods \citep{kerbl20233d} have emerged as the dominant approach. 
This advantage stems from two key factors: first, 3DGS employs explicit geometric representations that naturally facilitate dynamic modeling; second, its highly optimized CUDA rasterization pipeline eliminates the need for intensive sampling and querying of neural fields~\cite{es,evos,gs-survey}, significantly accelerating both rendering and training.



Recent 3DGS-based dynamic scene representations fall into two categories: 
1) \emph{Explicit} approaches, which extend the 3D Gaussian to 4D Gaussian primitives by adding a temporal dimension to approximate the spatiotemporal 4D volume of dynamic scenes~\citep{yang2023real,duan20244d}.
Although 4D Gaussians achieve higher visual quality and faster rendering speed, these works suffer from substantial storage requirements for numerous Gaussians and their 4D parameters as they cannot leverage inherent cross-spatiotemporal correlations.
2) \emph{Implicit} approaches, which employ consistent deformable 3D Gaussians as the underlying structure to characterize scenes, interpreting motion at each timestamp as a deformation of the underlying structure of different magnitudes~\citep{wu20244d4dgs,yang2023deformable,bae2024per}.
These deformation-based methods offer a more compact representation than explicit methods by leveraging intrinsic correlations across spatial and temporal conditions through deformations, enabling effective cross-spatiotemporal information sharing.
However, these implicit representations struggle to handle complex real-world motions and generally exhibit slower rendering speeds. 
%

Motivated by these challenges, we introduce SD-GS for dynamic scene reconstruction to address the balance between storage efficiency and the capability to model complex real-time motions. 
Our framework extends the anchor-based scaffold representation from static scenes to deformable 3D Gaussian-based dynamic scene reconstruction. Since local 3D Gaussians in spatiotemporal domains typically possess similar feature information, we propose to use deformable 3D structured anchors initialized from a sparse grid of SfM points as the underlying scene representation, achieving a more compact structure compared to explicit methods. The attributes of local neural 3D Gaussians can be predicted from anchor features adapted to various timestamps and viewing angles.

Furthermore, previous densification strategies in static scene reconstruction accumulated Gaussian gradients and applied growth in underfit regions based on predefined gradient thresholds. When applied to dynamic scene reconstruction, this approach tends to generate numerous redundant Gaussians in static regions, accompanied by increased deformation calculation overhead and lower FPS, while producing insufficient anchors in dynamic regions, leading to poor reconstruction quality in these areas. 
%
To overcome this limitation, we propose a deformation-aware densification strategy to achieve adaptive and efficient anchor allocation, directing new anchors toward poorly reconstructed high-dynamic regions instead of static scene parts.

In summary, our contributions are as follows: \textbf{1)} We introduce structured 3D Gaussians and a meticulously designed time-aware architecture to model dynamic scenes, significantly reducing the model size. \textbf{2)} We propose a deformation-aware densification method to further enhance the representation ability of the deformation grid in complex dynamics, which also reduces storage costs. \textbf{3)} Extensive experiments demonstrate that our approach significantly reduces storage requirements by 60\% on average and achieves 2x faster rendering speed, while maintaining or even exceeding state-of-the-art visual quality.

\section{Related Work}

\subsection{Dynamic 3D Gaussians}
Based on the formulation of the movement of the objects, the existing Dynamic 3D Gaussians can be divided into explicit and implicit methods based on the representation of time. The explicit methods~\cite{yang2023real,duan20244d,li2024spacetime} are built on the 4D Gaussians, with one more dimension representing the timestamp, which requires substantial memory for training and rendering. In contrast, the implicit methods ~\cite{wu20244d4dgs,bae2024per,yang2023deformable,luiten2024dynamic,huang2024sc} employ the deformation grid to model the movement, which greatly utilize the spatiotemporal correlations to reduce the memory and storage requirement. For instance, D-3DGS~\cite{luiten2024dynamic} models dynamic scenes by allowing the positions and rotation matrices of 3DGS to change over time. Deformable 3DGS~\cite{yang2023deformable} uses an MLP to model a deformation field based on time and the canonical Gaussian space. SC-GS~\cite{huang2024sc} bounds dense 3DGS with sparse control points, calculating the movement of Gaussians in a coarse-to-fine manner. However, these implicit methods ignore the inner redundancy of the canonical model, 3D Gaussians. Thus, we introduce a scaffold representation to replace the 3D Gaussians, which further reduces the memory requirement. Besides, we carefully design a deformation grid to enable the high-quality reconstruction for each timestamp.

Although contemporary works~\cite{cho2024scaffold4d,kwak2025modecgs} also utilize scaffold representations, they employ a different deformation strategy from us. For example, Scaffold4D~\cite{cho2024scaffold4d} still keeps the temporal dimension of the 4D Gaussians, while our method utilizes a memory-efficient deformation grid. 

\subsection{Gaussian Densification}
Gaussian densification plays a pivotal role in recovering accurate scene geometry. The vanilla 3DGS approach initializes sparse points from structure-from-motion (SfM)~\cite{schonberger2016structure} and employs adaptive density control~\cite{kerbl20233d}, first selecting Gaussians based on image-space gradients and then cloning or splitting them according to their scale. Intuitively, a well-designed densification strategy can enhance optimization convergence, accelerating the overall training process. Most existing methods, with limited exceptions such as~\cite{fang2024mini,kim2024color,mallick2024taming}, prioritize improving rendering quality, often at the expense of increased computational overhead.  Recent studies, such as~\cite{rota2024revising,kim2024color,mallick2024taming,ye2024absgs,zhang2024pixel}, have refined this strategy by incorporating image-space priors and intrinsic Gaussian properties for more informed selection. Other approaches integrate multi-view constraints~\cite{cheng2024gaussianpro,du2024mvgs,li2024mvg}, leverage advanced optimization techniques, or analyze point cloud geometry~\cite{fang2024mini}. However, densification strategies for dynamic Gaussians remain underexplored. In this work, we propose a deformation-aware densification strategy tailored to our dynamic Gaussian representation, which not only reduces storage consumption but also improves reconstruction quality for complex motion patterns.

\section{Preliminary}

\begin{figure*}[ht]
    \centering
    \includegraphics[width=1\textwidth]{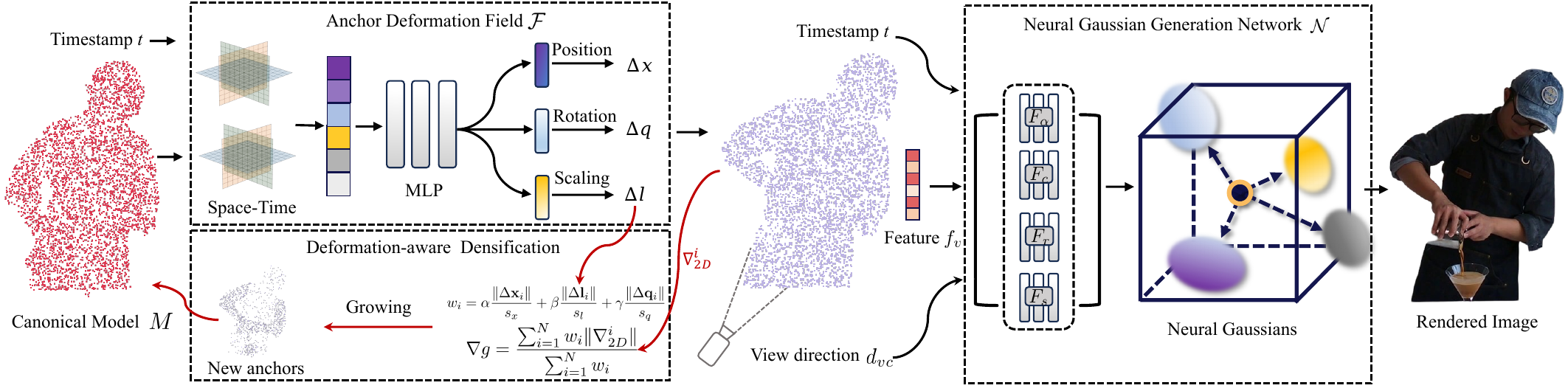}
    \caption{\textbf{Overview of SD-GS.} 
    We introduce the Canonical Gaussian Model $\mathcal{M}$ as the geometric structure of dynamic scenes. 
    Given the Canonical Gaussian Model $\mathcal{M}$ and timestamp $t$, the Anchor Deformation Field $\mathcal{F}$ 
    transforms the original Canonical Gaussian Model $\mathcal{M}$ into the Deformed Gaussian Model $\mathcal{M}'$. 
    Neural Gaussians at the specific timestamp are then generated through the Neural Gaussians Generation Network 
    $\mathcal{N}$. These neural Gaussians are subsequently splatted to produce rendered images using a 3D 
    Gaussian splatting pipeline. To better model complex real-world dynamics, we propose a deformation-aware densification strategy that 
    encourages new anchors to grow efficiently in under-reconstructed high-dynamic regions while reducing 
    redundancy in static areas.}
    \label{fig:overview}
\end{figure*}

\subsection{3D Gaussian Splatting}

{3DGS} \citep{kerbl20233d} is an explicit 3D representation in the form of point clouds, utilizing Gaussians to model the points. Each Gaussian is characterized by a covariance matrix $\mathbf{\Sigma}$ and a center point $\mu$, which is referred to as the mean value of the Gaussian: $G(x) = e^{-\frac{1}{2}(x-\mu)^\top \mathbf{\Sigma}^{-1}(x-\mu)}$.
To maintain the positive definiteness of the covariance matrix $\mathbf{\Sigma}$, 
3DGS decomposes $\mathbf{\Sigma}$ into a scaling matrix $\mathbf{S} = {\rm diag}(\mathbf{s}), \mathbf{s} \in \mathbb{R}^3$ and a rotation matrix $\mathbf{R}$: $\mathbf{\Sigma} = \mathbf{R}\mathbf{S}\mathbf{S}^\top \mathbf{R}^\top$.
The rotation matrix $\mathbf{R}$ is parameterized by a rotation quaternion $\mathbf{r} \in \mathbb{R}^4$. The backpropagation process is illustrated in \citep{kerbl20233d}.

When rendering novel views, the technique of splatting
\citep{ewa_volume_splatting,yifan2019differentiable} is employed for the Gaussians 
within the camera planes. As introduced by \citep{zwicker2001surface}, 
using a viewing transform denoted as $\mathbf{W}$ and 
the affine transform $\mathbf{J}$, the covariance
matrix $\mathbf{\Sigma}'$ in camera coordinates system can be computed by $\mathbf{\Sigma}' = \mathbf{J} \mathbf{W} \mathbf{\Sigma} \mathbf{W}^\top \mathbf{J}^\top$.
Specifically, for each pixel, the color and opacity of Gaussians are computed using $G(x)$. 
The blending of $N$ ordered points that overlap the pixel is given by: $C = \sum_{i\in N}{c_i \alpha_i \prod_{j=1}^{i-1}(1 - \alpha_j)}$. Here, $c_i$ and $\alpha_i$ represent the density and color of this point
computed by a Gaussian with covariance $\mathbf{\Sigma}$ multiplied by
a per-point opacity and SH color coefficients.

\subsection{Scaffold-GS}\label{sec:pre_sca}

{Scaffold-GS} \citep{lu2024scaffold} is a variant of 3DGS, widely adopted in 3DGS compression~\cite{hac2024,ren2024octree,wang2024contextgs,xie2024mesongs,xie2024sizegs} due to its low storage requirements. It introduces \textit{anchor points} to capture common attributes of local 3D Gaussians. Specifically, the \textit{anchor points} are initialized from neural Gaussians by voxelizing the 3D scenes. Each anchor point has a context feature $\rvf \in \sR^{32}$, a location $\rvx \in \sR^3$, a scaling factor $\rvl \in \sR^6$ and $k$ learnable offset $\rmO \in \sR^{k\times 3}$. Given a camera at $\rvx_c$, anchor points are used to predict the view-dependent neural Gaussians in their corresponding voxels as follows,
\begin{equation}
\{\rvc^i, \rvr^i, \rvs^i, \alpha^i\}_{i=0}^k = \text{MLP}(\rvf, \bm{\sigma}_{c}, \vec{\mathbf{d}}_{c}),
\end{equation}
where $\bm{\sigma}_{c} = ||\rvx-\rvx_c||_2$, $\vec{\mathbf{d}}_{c}=\frac{\rvx-\rvx_c}{||\rvx-\rvx_c||_2}$, the superscript $i$ represents the index of neural Gaussian in the voxel, $\rvs^i, \rvc^i \in \sR^3$ are the scaling and color respectively, and $\rvr^i \in \sR^4$ is the quaternion for rotation. 
The positions of neural Gaussians are then calculated as
\begin{equation}
    \{\bm{\mu}^0, ..., \bm{\mu}^{k-1}\} = \rvx + \{\rmO^0, ..., \rmO^{k-1}\}\cdot \rvl_{:3},
\end{equation}
where $\rvx$ is the learnable positions of the anchor and $\rvl_{:3}$ is the base scaling of its associated neural Gaussians. After decoding the properties of neural Gaussians from anchor points, the remaining steps are the same as 3DGS~\citep{kerbl20233d}. By predicting the properties of neural Gaussians from the anchor features and saving the properties of anchor points only, Scaffold-GS greatly eliminates the redundancy among 3D neural Gaussians and decreases the storage demand.

\subsection{Gaussian Deformation Field Network}

The Gaussian Deformation Field Network is a core component of 4DGS \citep{wu20244d4dgs} designed to model dynamic 3D scenes across space and time. It consists of a spatial-temporal structure encoder \(\mathcal{H}\) and a multi-head Gaussian deformation decoder \(\mathcal{D}\). Given a 3DGS model \(G\) and a timestamp \(t\), the network predicts deformations \(\Delta G = \mathcal{F}(G, t)\) to generate temporally coherent 4D Gaussians \(G' = G + \Delta G\).  

The encoder \(\mathcal{H}\) leverages a memory-efficient multi-resolution HexPlane decomposition \citep{cao2023hexplane,fridovich2023k_k-planes}, which projects 4D spatiotemporal features onto six 2D planes: \((x,y)\), \((x,z)\), \((y,z)\), \((x,t)\), \((y,t)\), and \((z,t)\). Each plane employs bilinear interpolation to aggregate multi-scale voxel features \(f_h\), followed by a lightweight MLP \(\phi_d\) to fuse these features into a unified deformation embedding \(f_d\). This hierarchical encoding captures localized spatial-temporal correlations among neighboring Gaussians while minimizing computational overhead.  

The decoder \(\mathcal{D}\) utilizes separate MLP heads to predict deformation parameters for position (\(\Delta \mathcal{X}\)), rotation (\(\Delta \rvr\)), and scaling (\(\Delta \rvs\)) as:  
\begin{equation}
(\Delta \mathcal{X}, \Delta \rvr, \Delta \rvs) = (\phi_x(\rvf_d), \phi_{\rvr}(\rvf_d), \phi_{\rvs}(\rvf_d)),
\end{equation}
yielding deformed Gaussians \(G' = \{\mathcal{X}+\Delta \mathcal{X},\, \rvr+\Delta \rvr,\, \rvs+\Delta \rvs,\, \sigma,\, C\}\). The framework preserves the differential splitting mechanism, enabling efficient novel view synthesis by rendering deformed Gaussians through \(G'\). This approach balances expressiveness and efficiency, making it suitable for dynamic scene modeling.

\section{Method}

\textbf{Overview.} We introduce a novel compact representation for dynamic scenes, which represents the scene using a set of deformable 3D Gaussians in an anchor-based framework. 
In this section, we will describe each component and its corresponding optimization process. 
In Sec.~\ref{sec-model-arch}, we introduce the overall framework, including the Canonical Gaussian Model $\mathcal{M}$, and the method of obtaining the Canonical Gaussian Model at a specific timestep via Anchor Deformation Field $\mathcal{F}$. Additionally, to model motion with finer details, we incorporate additional temporal information to supervise Neural Gaussians Generation Network $\mathcal{N}$. In Sec.~\ref{sec-deform-dense}, we elaborate on how our anchors efficiently grow in complex dynamic regions with reconstruction deficiencies, achieving better visual performance with fewer anchor points. Furthermore, the optimization framework will be introduced in Sec.~\ref{sec-loss}.

\subsection{Model Architecture}
\label{sec-model-arch}
The overview of our framework is illustrated in Fig.~\ref{fig:overview}, which consists of three main components: the Canonical Gaussian Model $\mathcal{M}$, the Anchor Deformation Fields $\mathcal{F}$, and the Neural Gaussian Generation Network $\mathcal{N}$. 

\vspace{1mm}\noindent\textbf{Canonical Gaussian Model $\mathcal{M}$.} To reduce memory requirements, we replace the 3D Gaussians with Scaffold-GS \cite{lu2024scaffold}, which significantly decreases memory usage through its anchor structure.
The Canonical Gaussian Model $\mathcal{M}$ represents the entire scene's geometric structure by combining anchor points with a set of local neural Gaussians. As mentioned in Sec.~\ref{sec:pre_sca}, each anchor point is characterized by a local context feature vector $\rvf_v \in \mathbb{R}^{32}$, a 3D position $\rvx \in \mathbb{R}^3$, a scaling factor $\rvl \in \mathbb{R}^6$, rotation quaternion $\rvq \in \mathbb{R}^4$ and $k$ learnable offsets $\rmO \in \mathbb{R}^{k \times 3}$. The last three dimensions of the scaling factor $\rvl$ enable anisotropic scaling of the neural Gaussians, while the first three dimensions, together with learnable offsets $\rmO$, determine the positions of $k$ neural Gaussians. The rotation quaternion $\rvq$ primarily influences the view frustum computation because we restrict the prediction of the neural Gaussian to the anchors within the view frustum during inference. To better learn high-quality anchor distributions in dynamic regions, we initialize the anchors using the sparse points from COLMAP \cite{schonberger2016structure}, and then train a static Canonical Gaussian Model without anchor deformation using all multi-view images from the dynamic video during the coarse stage.

\vspace{1mm}\noindent\textbf{Anchor Deformation Field $\mathcal{F}$.} 
Given the Canonical Gaussian Model $\mathcal{M}$ and timestamp $t$, the Anchor Deformation Field $\mathcal{F}$ transforms the original Canonical Gaussian Model $\mathcal{M}$ into Deformed Gaussian Model $\mathcal{M}' = \Delta\mathcal{M} + \mathcal{M}$. 
The deformation $\Delta\mathcal{M}$ is introduced by $\mathcal{F}(\mathcal{M}, t)$, where a spatial-temporal structure encoder $\mathcal{H}$ encodes both temporal and spatial features of anchors $\rvf_d = \mathcal{H}(\mathcal{M}, t)$, and a multi-head anchor deformation decoder $\mathcal{D}$ predicts the deformation $\Delta\mathcal{M} = \mathcal{D}(\rvf_d)$. The deformation $\Delta\mathcal{M} = \{\Delta \rvx, \Delta \rvl, \Delta \rvq\}$, where $\Delta \rvx$, $\Delta \rvl$, and $\Delta \rvq$ represent the deformation of anchor's 3D position, scaling factor, and rotation quaternion, respectively. These are computed by separate MLPs:
\begin{equation}
(\Delta \rvx, \Delta \rvl, \Delta \rvq) = (\phi_x(\rvf_d), \phi_l(\rvf_d), \phi_q(\rvf_d)).
\end{equation}
Our strategy quantifies the deformation of anchors, encouraging new anchors to grow in dynamic regions that lack reconstruction rather than static regions. The details of this strategy are further discussed in Sec.~\ref{sec-deform-dense}.

\vspace{1mm}\noindent\textbf{Neural Gaussians Generation Network $\mathcal{N}$.}
To enhance the model's temporal perception, we incorporate temporal information into Neural Gaussians Generation Network $\mathcal{N}$. Specifically, given the Deformed Gaussian Model $\mathcal{M}'$ under a certain moment, we determine the positions for $k$ neural Gaussians as follows:
\begin{equation}
\{\bm{\mu}_0, \ldots, \bm{\mu}_{k-1}\} = \rvx' + \{\rmO_0, \ldots, \rmO_{k-1}\} \cdot \rvl_{1:3}.
\end{equation}
where $\rvx'$ represents the deformed position of each visible anchor.
The attributes of each neural Gaussian (opacity $\alpha_i \in \mathbb{R}$, color $\rvc_i \in \mathbb{R}^3$, rotation $\rvr_i \in \mathbb{R}^4$ and scaling $\rvs_i \in \mathbb{R}^3$) are predicted through individual MLP decoders. Specifically, we construct a conditional vector $[\rvf_v, \rvd_{vc}, \phi(t)]$ by combining the anchor feature $\rvf_v$, viewing direction $\rvd_{vc}$, and temporal embedding $\phi(t)$. The conditional vector is fed into four independent fully-connected networks $F_\alpha$, $F_{\rvc}$, $F_{\rvr}$, $F_{\rvs}$, which decode all attributes of the neural Gaussians:


\begin{equation}
\begin{aligned}
&\left\{
\begin{array}{l}
\alpha_i = F_\alpha(\rvf_v, \rvd_{vc}, \phi(t)) \\
\rvc_i = F_{\rvc}(\rvf_v, \rvd_{vc}, \phi(t)) \\
\rvr_i = F_{\rvr}(\rvf_v, \rvd_{vc}, \phi(t)) \\
\rvs_i = F_{\rvs}(\rvf_v, \rvd_{vc}, \phi(t))
\end{array}
\right.
\end{aligned}
\end{equation}

After obtaining the 3D neural Gaussians at the given timestamp, we render them using the existing efficient 3D Gaussian splatting method \cite{kerbl20233d}, which applies to neural Gaussians within the view frustum and with opacity greater than a certain threshold.

\begin{figure*}[ht]
    \centering
    \includegraphics[width=1\textwidth]{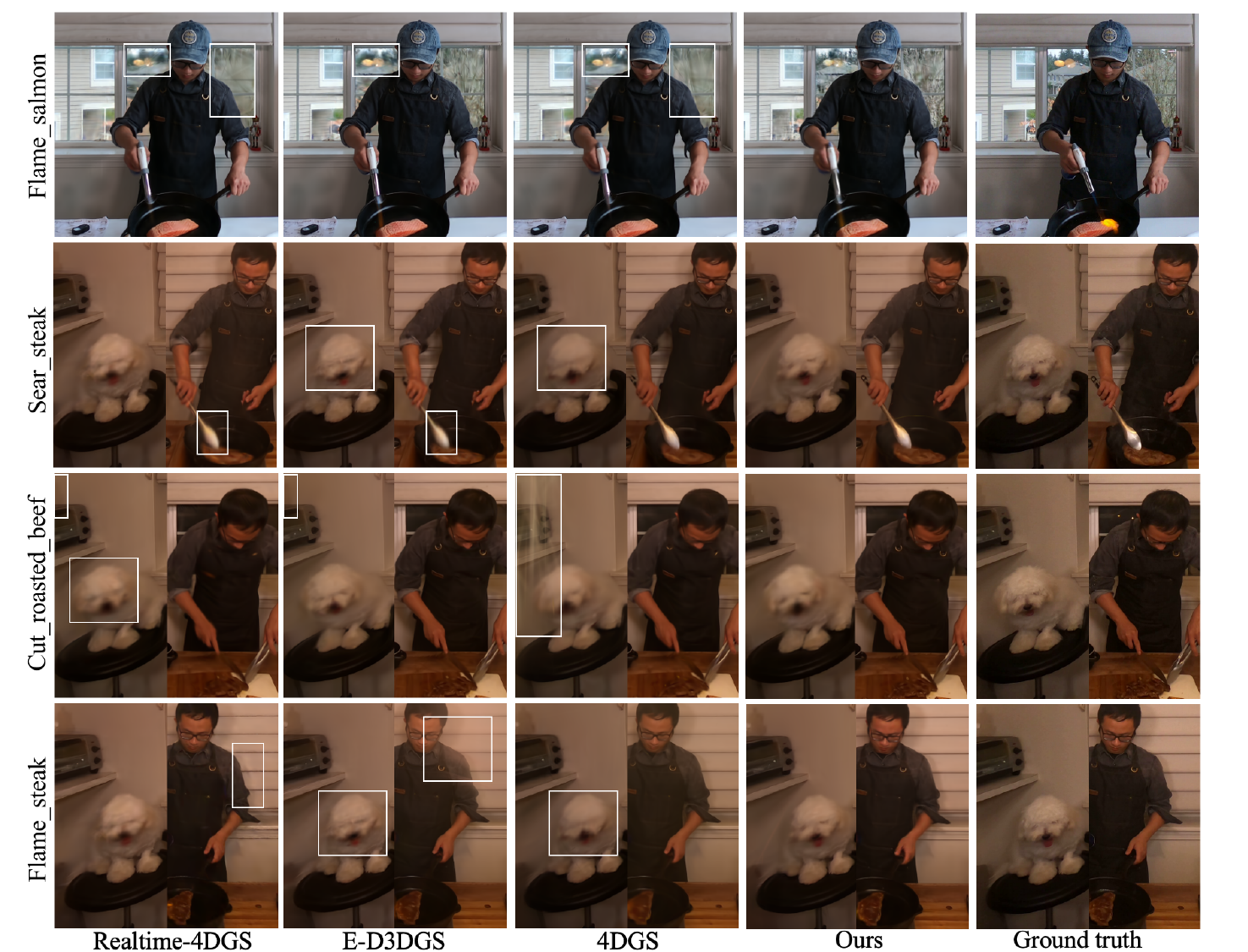}
    \caption{\textbf{Qualitative results on N3DV dataset.} The white boxes highlight under-reconstructed regions. Our method demonstrates superior fidelity across both dynamic and static areas of the scene.}
    \label{fig:N3DV}
\end{figure*}

\subsection{Deformation-aware Densification Strategy}
\label{sec-deform-dense}
The approach represents temporal variations through the deformation of 3D Gaussian tends to produce relatively lower visual quality in dynamic regions. The key factor to achieving high-quality results lies in efficiently growing new anchors in under-reconstructed dynamic regions while reducing redundant anchors in static areas.

Scaffold-GS \cite{lu2024scaffold} introduced an error-based anchor expansion strategy for static scenes, which grows new anchors where neural Gaussians find significant. This approach collects gradients of neural Gaussians by averaging over N iterations, denoted as $\nabla g$. Then, new anchors are placed in underfitting regions based on predefined gradient thresholds. However, when directly applied to dynamic scene reconstruction, the transient motions occurring in dynamic regions, due to their short duration, fail to acquire sufficient $\nabla g$ to generate anchors for motion modeling, as they are inevitably penalized by the denominator N, regardless of their actual errors. Consequently, this method fails to achieve satisfactory visual results.

To address the issue, we propose a deformation-aware densification strategy that quantifies anchor deformations and assigns larger gradient weights to active neural Gaussians. This enables anchors to adaptively grow based on motion dynamics. We formulate $\nabla g$ as:
\begin{equation}
\nabla g = \frac{\sum_{i=1}^{N} w_i \|\nabla_{2D}^i\|}{\sum_{i=1}^{N} w_i}, 
\end{equation}
where $\nabla_{2D}^i$ is the 2D positional gradient of neural Gaussians in the i-th iteration, and the weight term $w_i$ is determined by the anchor's deformation magnitude as follows:
\begin{equation}
\label{eq:wi}
w_i = \alpha\frac{\|\Delta \mathbf{x}_i\|}{s_x} + \beta\frac{\|\Delta \mathbf{l}_i\|}{s_l} + \gamma\frac{\|\Delta \mathbf{q}_i\|}{s_q}, 
\end{equation}
where $\Delta \mathbf{x}_i$, $\Delta \mathbf{l}_i$, and $\Delta \mathbf{q}_i$ represent the anchor's deformation in position, scaling, and rotation, respectively. $s_x$, $s_l$ and $s_q$ are normalization reference values for each deformation magnitude. To effectively highlight anchor points with significant deformation, we use the 90th percentile value of each deformation type in each iteration as the normalization reference values. The weighting coefficients $\alpha$, $\beta$, and $\gamma$ control the relative contribution of each deformation component.
Specifically, $\|\Delta \mathbf{x}_i\|$ and $\|\Delta \mathbf{l}_i\|$ are computed as the Euclidean distance between the anchor positions and scaling parameters before and after deformation.
The rotation deformation amplitude  $\|\Delta \mathbf{q}_i\|$ is:
\begin{equation}
\|\Delta \mathbf{q}_i\| = 2 \cdot \arccos \left( {\rm clip} \left( \left| \sum_{k=1}^4 \mathbf{q}_{\text{orig},i}^{(k)} \cdot \mathbf{q}_{\text{def},i}^{(k)} \right|, 0.0, 1.0 \right) \right), 
\end{equation}
where $\mathbf{q}_{\text{orig}}, \mathbf{q}_{\text{def}} \in \mathbb{R}^{N \times 4}$ represent the unit quaternions before and after anchor deformation. This formula calculates the absolute value of the quaternion dot product to measure rotation difference, where the absolute value operation ensures that the direction of rotation remains unchanged. The dot product result is clipped to the [0,1] range to avoid floating-point errors causing numerical instability. Then the $2 \cdot \arccos(\cdot)$ function maps this to the rotation angle ranging from [0, $\pi$].

Our strategy precisely identifies anchor points in dynamic regions with significant deformation, enabling these anchors to receive greater gradient weight rewards. This mechanism encourages the growth of new anchors in under-reconstructed dynamic areas rather than static backgrounds, thereby optimizing the anchor allocation mechanism to achieve adaptive spatial deployment of anchors and improve the efficiency of dynamic scene reconstruction. 

\subsection{Optimization Framework}
\label{sec-loss}
\textbf{Loss Design.} We select $\mathcal{L}_1$ loss over rendered pixel colors, SSIM loss $\mathcal{L}_{\text{SSIM}}$, a grid-based spatiotemporal total-variation loss $\mathcal{L}_{\text{tv}}$ \cite{cao2023hexplane, sun2022direct, fang2022fast} and volume regularization $\mathcal{L}_{\text{vol}}$ \cite{lombardi2021mixture}. The learnable parameters of anchors and MLPs are co-optimized by minimizing the rendering discrepancy. The entire training loss function is as follows:
\begin{equation}
\mathcal{L} = \mathcal{L}_1 + \lambda_{\text{SSIM}} \mathcal{L}_{\text{SSIM}} + \lambda_{\text{tv}} \mathcal{L}_{\text{tv}} + \lambda_{\text{vol}} \mathcal{L}_{\text{vol}},
\end{equation}
where the weighting coefficients $\lambda_{\text{SSIM}} = 0.2$, $\lambda_{\text{tv}} = 0.01$, and $\lambda_{\text{vol}} = 0.01$.

\section{Experiment}
\begin{figure*}[ht]
    \centering
    \includegraphics[width=0.8\textwidth]{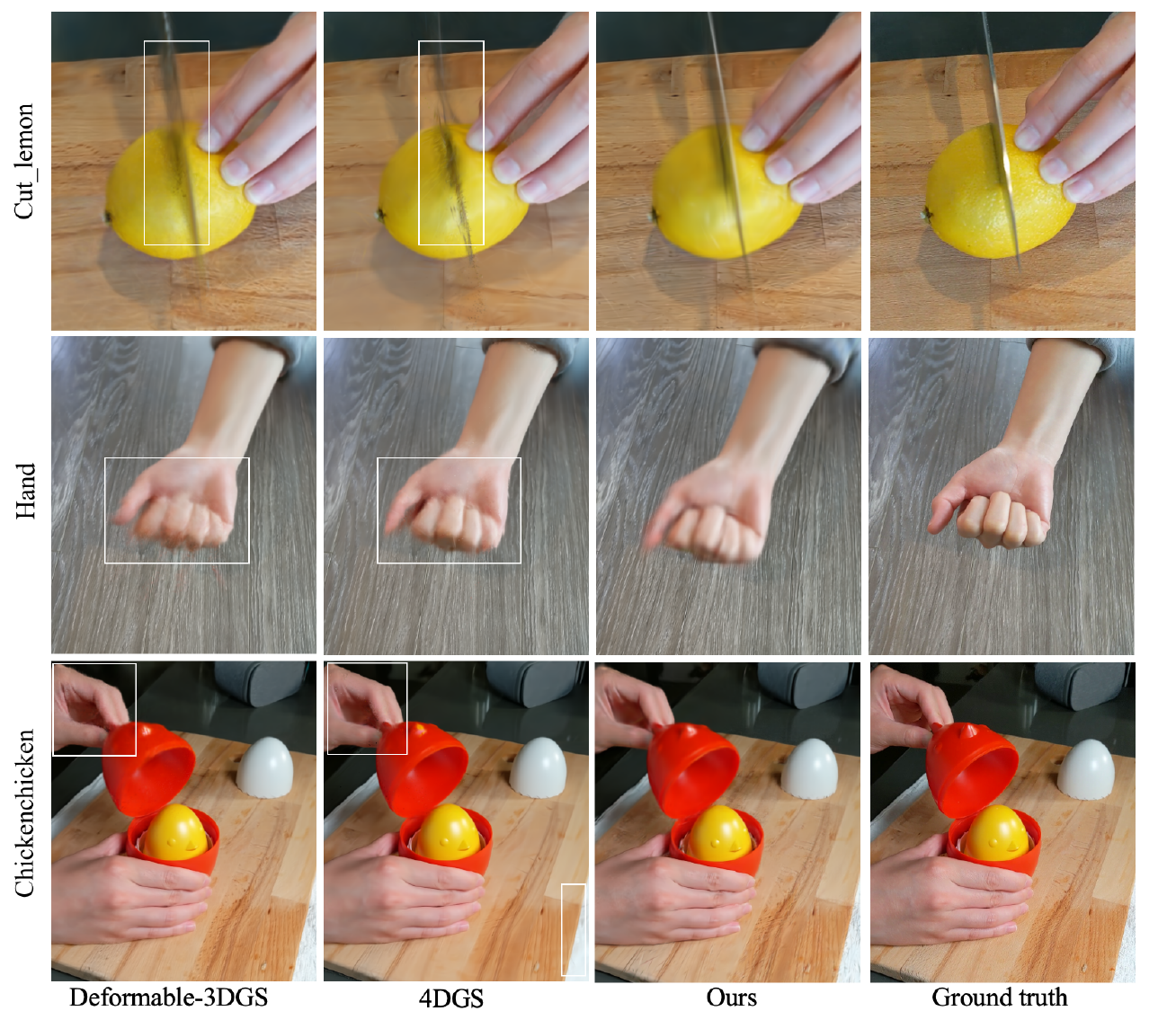}
    \caption{\textbf{Qualitative results on HyperNeRF dataset.} This figure presents qualitative comparisons on the HyperNeRF dataset. The white boxes highlight under-reconstructed regions. Our method demonstrates superior fidelity across both dynamic and static areas of the scene.}
    \label{fig:HyperNeRF}
\end{figure*}


\begin{table}[t]
\centering
\caption{\textbf{Quantitative results on N3DV dataset.} We computed the average metrics across all six scenes. The \colorbox{pink}{best} and the \colorbox{yellow}{second best} results are denoted by pink and yellow. } 
\label{tab:n3dv_results} 
\resizebox{\linewidth}{!}{
\begin{tabular}{@{}l|ccc|ccc@{}}
\hline
Model & \multicolumn{3}{c|}{Metrics} & \multicolumn{3}{c}{Computational cost} \\
\cline{2-7}
& PSNR$\uparrow$ & SSIM$\uparrow$ & LPIPS$\downarrow$ & Training time$\downarrow$ & FPS$\uparrow$ & Storage$\downarrow$ \\
\hline
4DGS \cite{wu20244d4dgs} & 30.94 & 0.936 & 0.056 & \colorbox{pink}{45min} & 34.7 & \colorbox{yellow}{59MB} \\
E-D3DGS \cite{bae2024per} & 30.86 & 0.938 & \colorbox{yellow}{0.048} & 3h20min & 45.4 & 64MB \\
Realtime-4DGS \cite{yang2023real} & \colorbox{yellow}{31.11} & \colorbox{yellow}{0.939} & 0.050 & 7h50min & \colorbox{yellow}{45.9} & 7970 MB \\
\textbf{Ours} & \colorbox{pink}{31.35} & \colorbox{pink}{0.942} & \colorbox{pink}{0.047} & \colorbox{yellow}{80min} & \colorbox{pink}{82.1} & \colorbox{pink}{22MB} \\
\hline
\end{tabular}
}
\end{table}

\begin{table}[t]
\centering
\caption{\textbf{Quantitative results on HyperNeRF dataset.} The \colorbox{pink}{best} and the \colorbox{yellow}{second best} results are denoted by pink and yellow. 
\label{tab:Hypernerf_results} 
\textsuperscript{1} uses the metric from the original paper.} 
\resizebox{\linewidth}{!}{
\begin{tabular}{@{}l|ccc|ccc@{}}
\hline
Model & \multicolumn{3}{c|}{Metrics} & \multicolumn{3}{c}{Computational cost} \\
\cline{2-7}
& PSNR$\uparrow$ & SSIM$\uparrow$ & LPIPS$\downarrow$ & Training time$\downarrow$ & FPS$\uparrow$ & Storage$\downarrow$ \\
\hline
4DGS \cite{wu20244d4dgs} & \colorbox{yellow}{25.72} & \colorbox{pink}{0.744} & \colorbox{yellow}{0.230} & \colorbox{yellow}{30min} & 34.1 & 74MB \\
E-D3DGS \cite{bae2024per}\textsuperscript{1} & 25.43 & 0.697 & 0.231 & 2h5min & \colorbox{yellow}{73.4} & \colorbox{yellow}{50MB} \\
Deformable 3DGS \cite{yang2023deformable} & 24.91 & 0.705 & 0.243 & 1h30min & 14.2 & 172 MB \\
\textbf{Ours} & \colorbox{pink}{25.79} & \colorbox{yellow}{0.737} & \colorbox{pink}{0.221} & \colorbox{pink}{28min} & \colorbox{pink}{79.7} & \colorbox{pink}{43MB} \\
\hline
\end{tabular}
}
\end{table}

In Sec.~\ref{sec:Settings}, we introduce the datasets, metrics, and baselines. In Sec.~\ref{sec:Results}, we present the performance of our method on two different datasets and compare it with state-of-the-art methods based on 3D Gaussian Splatting. Subsequently, in Sec.~\ref{sec:Ablation}, we conduct ablation studies to demonstrate the effectiveness of each module.

\subsection{Experimental Settings}
\label{sec:Settings}

\vspace{1mm}\noindent\textbf{Datasets.} We evaluate our method on two representative real-world dynamic scene datasets: \textbf{1) Neural 3D Video dataset (N3DV)~\cite{li2022neural3d}} contains 6 real-world scenes. These scenes feature relatively long durations and diverse motions, some containing multiple moving objects. Each scene has approximately 20 synchronized videos. Except for the \textit{flame\_salmon} scene, which consists of 1200 frames, all other scenes comprise 300 frames. For each scene, we select one camera view for testing while using the remaining views for training. \textbf{2) HyperNeRF~\cite{park2021hypernerf}} is captured with 1-2 cameras, following straightforward camera motion. It contains complex dynamic variations, such as human movements and object deformations.



\vspace{1mm}\noindent\textbf{Baselines.} We compare our method against several state-of-the-art works in dynamic scene reconstruction, including deformation-based methods like 4DGS \cite{wu20244d4dgs} and Deformable 3DGS \cite{yang2023deformable}, as well as 4D Gaussian-based methods like Real-time 4DGS \cite{yang2023real}. We utilized their official code to test their performance.

\vspace{1mm}\noindent\textbf{Metrics.} To evaluate reconstruction quality, we employ peak-signal-to-noise ratio (PSNR), structural similarity index (SSIM) \cite{wang2004image}, and perceptual quality measure LPIPS \cite{zhang2018unreasonable} with an AlexNet Backbone to assess the rendered images. Additionally, we evaluate storage efficiency by calculating the output file size as storage (MB), including point cloud files, MLP weights, and other relevant components. We also measured the rendering speed (FPS).

\vspace{1mm}\noindent\textbf{Implementation Details.} To provide better anchor initialization, we first warm up by training a static ScaffoldGS without any deformation for 3000 iterations during the coarse stage. Subsequently, in the fine stage of 140k iterations, we train the anchor deformation field network along with the learnable parameters of the anchors. 
Our implementation is based on the PyTorch \cite{paszke2019pytorch} framework and runs on a single NVIDIA RTX 3090 GPU. The optimization parameters of the entire framework are appropriately fine-tuned with reference to Scaffold-GS \cite{lu2024scaffold} and 4DGS \cite{wu20244d4dgs}. We set $\alpha=0.8$, $\beta=0.1$, and $\gamma=0.1$ for Eq.~\ref{eq:wi}.

\subsection{Results Analysis}
\label{sec:Results}

\vspace{1mm}\noindent\textbf{N3DV.} We deliver quantitative results on the N3DV dataset in Table~\ref{tab:n3dv_results}. While deformation-based methods ~\cite{wu20244d4dgs, bae2024per} offer compact memory usage, their FPS performance is significantly constrained due to the computational overhead of deformation calculations for numerous Gaussians. In contrast, our method not only delivers superior image quality but also achieves approximately 100\% higher FPS while reducing storage costs by around 65\%.
Notably, compared to the 4D Gaussian-based approach~\cite{yang2023real}, our method achieves a remarkable 362× reduction in storage requirements and 6.4× faster training time while simultaneously improving FPS by approximately 82\%. Overall, our method outperforms state-of-the-art Gaussian Splatting-based methods in both visual quality and rendering efficiency while maintaining an exceptionally compact model size.

To further evaluate the model performance, we provide qualitative comparisons with baselines in~\ref{fig:N3DV}. As highlighted in the boxed areas, existing methods often introduce artifacts and blurriness, particularly struggling with dynamic region reconstruction. In contrast, our method generates sharp and high-fidelity rendering results in both static and dynamic regions.

\vspace{1mm}\noindent\textbf{HyperNeRF.} 
Table~\ref{tab:Hypernerf_results} presents the quantitative results on the HyperNeRF dataset. The results demonstrate that our method achieves highly competitive reconstruction performance while achieving the fastest rendering speed, shortest training time, and minimal storage requirements.
Figure~\ref{fig:HyperNeRF} presents qualitative comparisons with Gaussian-based methods on the HyperNeRF dataset. Previous methods struggle to reconstruct regions with rapid motion, often producing blurry artifacts in dynamic areas, such as the moving hands and knife in the boxed areas. In contrast, our method achieves high visual fidelity in both static and dynamic regions, effectively mitigating motion-related distortions.

\subsection{Ablation Studies}\label{sec:Ablation}

\begin{table}[t]
\centering
\caption{\textbf{Ablation studies on each component of our method.} ``DAD" refers to deformation-aware densification strategy. ``TIN" refers to temporal injection in Neural Gaussians Generation Network $\mathcal{N}$. Experiments are conducted on the \textit{flame\_steak} scene of the Neural 3D Video dataset.}
\setlength{\tabcolsep}{0.5mm}  
\begin{tabular}{@{}l|ccc|cc@{}}
\hline
Model & PSNR$\uparrow$ & SSIM$\uparrow$ & LPIPS$\downarrow$ & Anchors$\downarrow$ & Storage$\downarrow$ \\
\hline
Ours w/o DAD & 32.61 & 0.951 & 0.037 & 53K & 26.28 \\
Ours w/o TIN & 31.15 & 0.949 & 0.041 & 40K & 22.13 \\
Ours w/o $\Delta x$ & 30.20 & 0.947 & 0.042 & 41K & 22.47 \\
Ours w/o $\Delta l$ & 32.43 & 0.955 & 0.038 & 41K & 22.55 \\
Ours w/o $\Delta q$ & 32.70 & 0.956 & 0.036 & 41K & 22.53 \\
Ours & 33.09 & 0.957 & 0.036 & 41K & 22.54\\
\hline
\end{tabular}
\label{tab:ablation}
\end{table}

\begin{figure}[t]
    \centering
    \includegraphics[width=1\columnwidth]{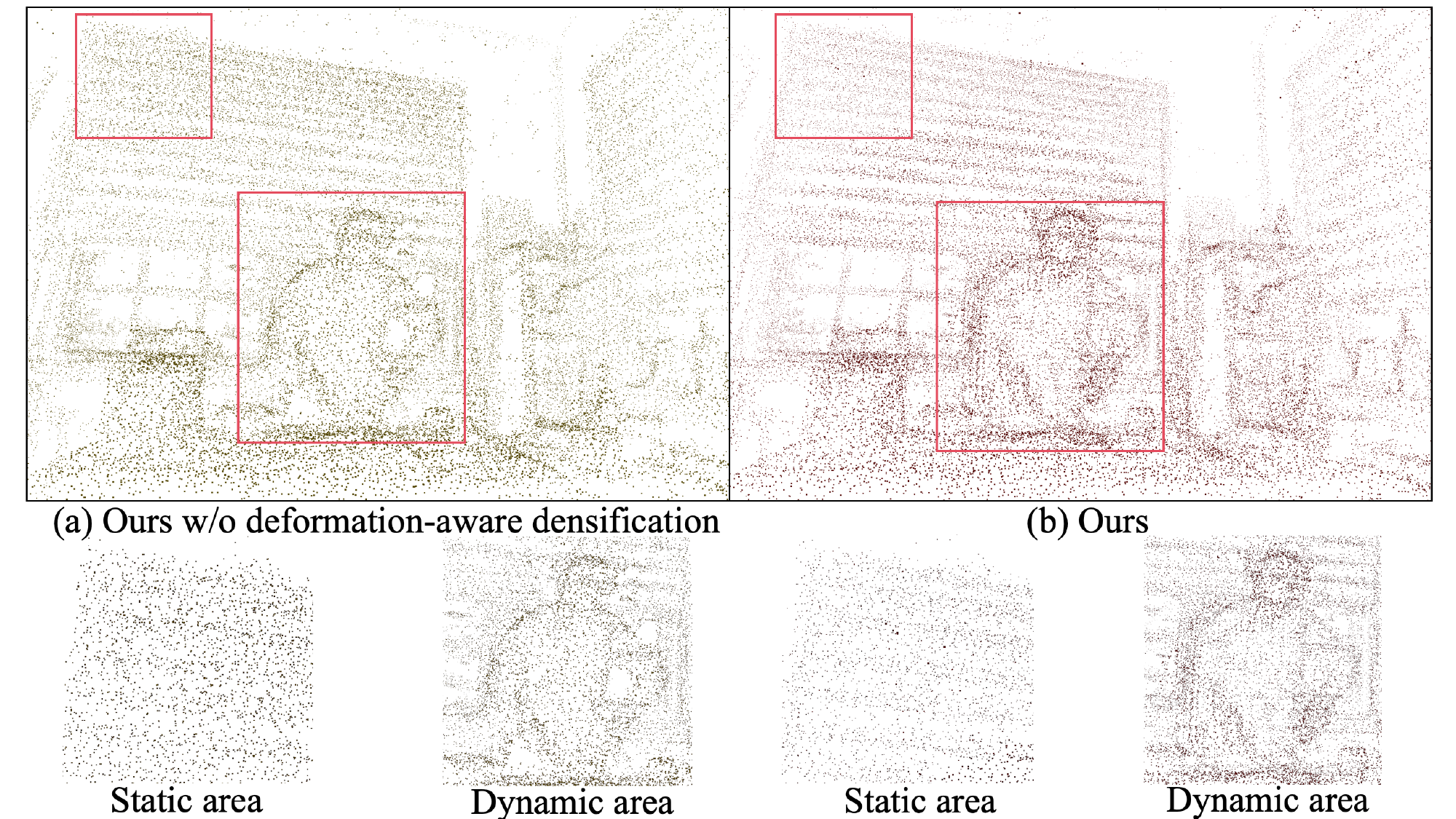}
    \caption{\textbf{Effectiveness of deform-aware densification strategy.} Without this strategy, static and dynamic regions exhibit similar anchor density, as shown in (a). By adopting the deform-aware densification strategy, the anchor distribution in dynamic regions becomes significantly denser compared to the static regions, as illustrated in (b).}
    \label{fig:anchors}
    \vspace{-5pt}
\end{figure}

\vspace{1mm}\noindent\textbf{Effects of deformation aware densification strategy.}
To demonstrate the effectiveness of our proposed deformation aware densification strategy, we conduct a comparative analysis of anchors spatial distribution. As illustrated in Fig.~\ref{fig:anchors}, traditional densification approaches maintained similar anchor point densities across both static backgrounds and dynamic regions. In contrast, our method selectively increases anchor density in under-reconstructed dynamic areas, enhancing motion representation while minimizing redundant anchors in static regions. This adaptive allocation improves the modeling of complex motion while reducing computational overhead in static areas.

\vspace{1mm}\noindent\textbf{Understanding the temporal injection in Neural Gaussians Generation Network $\mathcal{N}$.}
To achieve more refined motion modeling, we incorporate temporal information not only into the Anchor Deformation Field $\mathcal{F}$ but also into the MLPs within the Neural Gaussians Generation Network  $\mathcal{N}$. This temporal injection in $\mathcal{N}$ enables the generation of neural gaussians with time-varying properties, allowing our model to capture dynamic scene characteristics more effectively. As demonstrated in Table~\ref{tab:ablation}, the models with dual temporal information injection outperform the model without temporal injection in Neural Gaussians Generation Network $\mathcal{N}$.

\vspace{1mm}\noindent\textbf{Analysis of the deformation of each part.}
We introduce separate MLPs ($\phi_x$, $\phi_l$, $\phi_q$) in the Anchor Deformation Field $\mathcal{F}$ to model the temporal changes of anchors, including 3D position, scaling factor, and rotation quaternion. As shown in Table~\ref{tab:ablation}, anchor movement plays the most critical role in fitting dynamic scenes. The scaling factor primarily modulates the anisotropic scaling of neural Gaussians and the scaling of offsets, serving to simulate the stretching and twisting of surfaces at the microscopic level during non-rigid motion processes. The rotation quaternion of anchors mainly determines the visibility of anchors in the view frustum, which is used to model changes in anchor visibility during movement.

\section{Conclusion}
This paper introduces SD-GS, an innovative and compact framework designed to represent dynamic scenes. 
It achieves high visual quality while significantly reducing storage costs and improving FPS.
Additionally, our densification strategy effectively optimizes anchor point distribution in dynamic scenes to address the challenges of complex motion reconstruction. Extensive experiments demonstrate that our model achieves competitive reconstruction quality on challenging real-world dynamic scene datasets while substantially reducing model size.

{
    \small
    \bibliographystyle{ieeenat_fullname}
    \bibliography{main}
}

\end{document}